# A Value Driven Framework for Cybersecurity Innovation in Transportation & Infrastructure

Lampis Alevizos
*Volvo Group &
UCLan - School of Engineering and Computer Science*
lampis@redisni.org

Lalit Bhakuni
Volvo Group
lalit.bhakuni@outlook.com

Stefan Jäschke
Volvo Group
stefan.jaschke@pm.me

*Abstract —* **This paper introduces a value-driven cybersecurity innovation framework for the transportation and infrastructure sectors, as opposed to the traditional market-centric approaches that have dominated the field. Recontextualizing innovation categories into sustaining, incremental, disruptive, and transformative, we aim to foster a culture of self-innovation within organizations, enabling a strategic focus on cybersecurity measures that directly contribute to business value and strategic goals. This approach enhances operational effectiveness and efficiency of cyber defences primarily, while also aligns cybersecurity initiatives with mission-critical objectives. We detail a practical method for evaluating the business value of cybersecurity innovations and present a pragmatic approach for organizations to funnel innovative ideas in a structured and repeatable manner. The framework is designed to reinforce cybersecurity capabilities against an evolving cyber threat landscape while maintaining infrastructural integrity. Shifting the focus from general market appeal to sector-specific needs, our framework provides cybersecurity leaders with the strategic cyber-foresight necessary for prioritizing impactful initiatives, thereby making cybersecurity a core business enabler rather than a burden.**

## I. INTRODUCTION

Cybersecurity in transportation and infrastructure sectors nowadays is crucial for the operational integrity of systems in modern society. It not simply about protecting asset confidentiality, integrity, and availability (CIA), rather it goes beyond that [1]. These sectors are increasingly become interconnected and even reliant on digital technologies, facing rapidly expanding cyber threats. Therefore, a re-evaluation of the innovation strategies employed to protect such critical infrastructure becomes imperative [1].

Historically, cybersecurity innovation has largely been influenced by market-driven forces, often highlighting the development of solutions with broad commercial applicability. Although such an approach has undeniably led to significant technological advances, it does not always align with the specific needs and value propositions in the cybersecurity domain and within self-innovating organizations in the transportation and infrastructure sectors. These industries require a framework for cybersecurity innovation that prioritizes operational continuity, safety, security, and public trust over general market appeal. Oftentimes the cybersecurity innovation in these sectors, originates from within, knowing the details, various aspects, and unique challenges of the business itself [3].

To address this disparity, this paper introduces a business value-driven framework for cybersecurity innovation and cyber-foresight tailored to the unique demands of the transportation and infrastructure sectors. Building upon the established categories of disruptive, transformative, sustaining, and incremental innovations, we recontextualize them to reflect their contribution to the sectors' business values, such as efficiency, effectiveness, and the capacity to foster a culture of innovation within teams. By doing so, we aim to redirect the focus from market-driven outcomes to innovations that deliver tangible value to the organization's core functions. Innovators often concentrate on fostering interactions between ideas and talents without a long-term vision [4]. In contrast, we extend our focus beyond mere idea generation, emphasizing not only the importance of a prolonged, collaborative journey but also on generating tangible value throughout the process.

The objective of this paper is to critically analyse the traditional market-centric model of innovation and propose an alternative framework that underscores the direct benefits to business operations, especially in sectors where the stakes of cybersecurity breaches are particularly high. To achieve this, we research the theoretical foundations of such an approach, provide a pragmatic methodology to operationalize it within organizations, and discuss its broader implications for strategic decision-making in cybersecurity.

The structure of the paper is as follows: we begin by providing a background and critically discussing the limitations of cybersecurity in the transportation and infrastructure sectors. Next, we detail the theoretical rationale for a value-driven innovation framework, and finally, discuss the practical considerations for its implementation. Thereby, we contribute to the discourse on cybersecurity strategy and offer a pragmatic roadmap for organizations to enhance their defensive capabilities through innovative ways.

## II. Background & Literature Review

The strategic imperative of cybersecurity in the transportation and infrastructure sectors has been well-documented, with scholars and practitioners acknowledging the increasing sophistication of threats and the need for resilient defence mechanisms [1] [2] [3]. Within this context, the transportation and infrastructure sectors face unique challenges, such as the requirement to maintain uninterrupted services and the management of large, complex systems that are often part of the critical national infrastructure (CNI) [4].

The conventional innovation frameworks, while valuable in promoting technological advancement, have been critiqued for their limited scope in addressing the nuanced needs of critical infrastructure [5]. More specifically, the transportation and infrastructure sectors require a focused approach that integrates risk management and operational continuity at its core [6]. This review identifies a gap in current literature where the business value, precisely in terms of operational efficiency and effectiveness, is insufficiently linked with the types of innovation in cybersecurity [7].

Chesbrough et al. [8] introduced the open innovation framework, where collaboration with external partners can bring fresh perspectives and specialized knowledge to internal cyber defence strategies, enhancing the company's ability to respond to evolving threats. However, sharing sensitive information externally can pose security risks, and the focus might divert from internal process optimization, which is crucial for efficiency and effectiveness. Kim's blue ocean strategy [9] encourages creative thinking in identifying unique approaches to cyber defence, potentially leading to more effective internal solutions without direct market competition. Nonetheless, the primary aim of creating market spaces may not align perfectly with internal innovation focused on enhancing current cyber defence capabilities, thus the proposed value driven approach may be more suitable in this context. Debruyne's [10] customer-centric innovation framework could potentially align closely with the needs of internal users (as 'customers'), subject to repurposing. However, the framework may not fully address the strategic and overarching goals of the organization's cyber defence posture, focusing more on individual user needs. Moreover, the design thinking framework [11] brings a human-centred approach that allows cyber defence solutions to be tailored to the needs of internal stakeholders, but on the other hand, the iterative, empathetic process of design thinking might be time-consuming, resulting in paralysis by analysis phenomena and potentially clashing with the need for rapid implementation in demanding environments. The disruptive innovation framework by Christensen et al. [12] sets the groundwork for potential introduction of new internal technologies or practices that revolutionize a company's cyber defence approach. Nevertheless, disruptive innovations in the cyber defence context within an organization, will necessitate significant adjustment periods. Consequently, this could indicate a time lag between detecting a breach and responding to it. Furthermore, there is a growing recognition of the role that sustaining innovation plays in creating an innovative culture within organizations [13]. Therefore, by empowering teams with the responsibility for continual, iterative improvement, sustaining innovation can serve as a catalyst for more ground-breaking initiatives within the cybersecurity domain.

To synthesize the existing body of work, this paper draws upon original theories of innovation by Christensen et al. [12] and Professor Schumpeter's theory of innovation [14], while also incorporating contemporary insights from cybersecurity and industry-specific sources [15]. This work forges a framework that addresses the gaps identified, while also aligns with the strategic imperatives of the transportation and infrastructure sectors. Ultimately, the evolving cyber landscape and the rise of self-innovating organizations highlights a shift from traditional, market-driven innovation models, which often prioritize scalability and profitability over sector-specific needs.

## III. Framework

This section introduces a framework for cybersecurity innovation, with a unique focus on maximizing business value rather than conventional market-driven metrics. Our proposed framework is based in the principle that the true measure of innovation in cybersecurity lies in its capacity to enhance cyber resilience, mitigate risks, align closely with the strategic objectives of the organizations, improved efficiency, and effectiveness of cyber defences, and provide cyber-foresight. This is contrary to the traditional models that often prioritize market reach, commercialization, and financial profit.

Cyber-foresight is a strategic capability that enables organizations to anticipate, identify, and prepare for future cybersecurity threats and technological trends before those are established [16]. Cyber-foresight is at the core of our value-driven framework. It empowers organizations to proactively identify emergence phenomena and trends in the cyber space, hence cybersecurity strategies can balance between reactive and proactive, in addition to being predictive. This anticipatory stance is imperative in achieving cyber resilience and can even provide a competitive advantage in the transportation and infrastructure sectors.

Therefore, reorienting the innovation process towards these parameters, we present a more relevant and impactful four-stage approach for organizations seeking to drive internal innovation while facing the cyber threat landscape. The four stages defined as: (1) assign innovation category & establish organizational ownership, (2) cybersecurity innovation value proposition scoring, (3) balance resources and risks, (4) execution and value realization. The four stages, along with their respective activities, are illustrated in figure 1, which is further elaborated upon in this section. These four stages are governed by a cybersecurity innovation forum composed of members with diverse expertise.

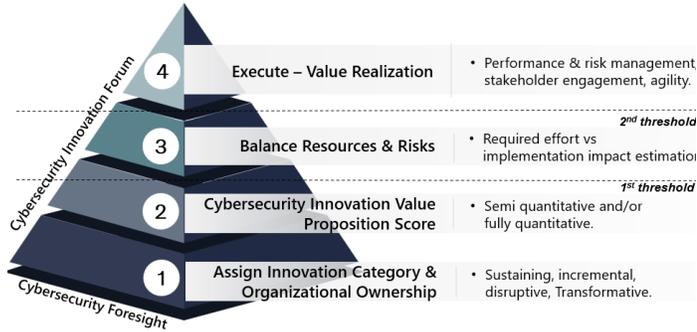

Figure 1 - Value-driven cybersecurity innovation framework.

## 1. Innovation Categorization

The framework details four distinct but complementary aspects of innovation in cybersecurity for the transportation and infrastructure sectors, namely: sustaining, incremental, disruptive, and transformative, visualized in figure 2.

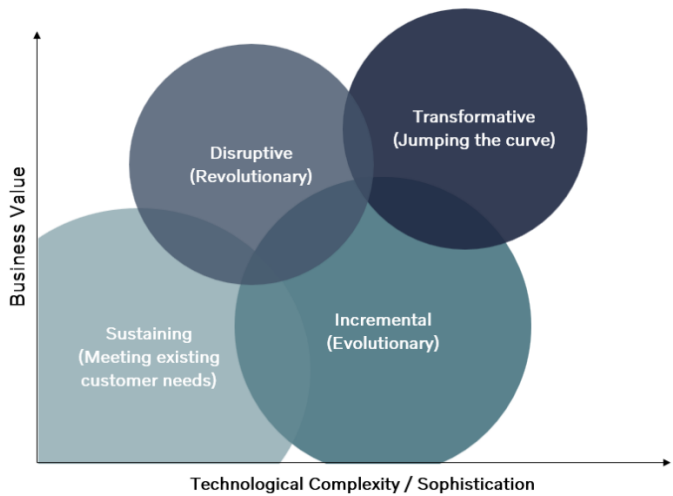

Figure 2 - Value driven innovation aspects.

Sustaining innovation focuses on refining and enhancing existing processes, while incremental innovation addresses evolving needs through minor yet impactful improvements. Disruptive innovation, on the other hand, introduces radical changes that reshape the cybersecurity landscape of organizations, and transformative innovation leads to fundamental shifts in practices and technologies. The ultimate goal of this framework is to cultivate a culture of self-innovation within organizational teams. Empowering teams to autonomously drive sustaining and incremental innovations, either independently or with support from a dedication cybersecurity innovation capability that provides tools and guidance, organizations can set the groundwork for innovation at all levels. This strategic approach allows dedicated innovation teams to concentrate their efforts on generating more disruptive and transformative innovations, thereby organizations can achieve a balanced and dynamic innovation ecosystem that responds to current and emerging cyber security challenges.

It is worth noting nonetheless, that excessive focus on any single category within the innovation portfolio can lead to challenges in implementation or diminish the overall impact of the initiative. For instance, a prevalence of incremental ideas might result in diminishing returns, thereby reducing the initiative's relevance over time [3]. Conversely, an abundance of disruptive innovations could present significant integration challenges due to the potential for widespread disruption they entail. We empirically estimate that the optimal composition of an innovation portfolio in the transportation and infrastructure sectors, considering their unique characteristics and as a general framework, a distribution consisting of 45% sustaining, 40% incremental, 10% disruptive, and 5% transformative innovation could serve as an effective initial allocation. Nonetheless this is entirely up to each organization's cultural dynamics, and the maturity of its innovation processes.

### a. Sustaining Innovations

Sustaining innovation in cybersecurity, particularly within the transportation and infrastructure sectors, is focused on meeting or even anticipating customer needs [20]. This aspect of innovation is about enhancing the effectiveness and efficiency of existing processes and capabilities of an organization, ultimately focused on extending their business value. Such innovations often arise from systematic efforts like hypothesis testing or thorough intellectual dialogues. For instance, helping stakeholders with innovative technologies or methodologies to understand and address their specific security needs. As a result, this may require developing of a structured and repeatable security process. It may also require developing a unique security solution, which addresses the unique challenge presented by such technologies. The value here lies in increasing efficiency through effective client interfacing. Thus, making sure business needs are understood and met, and providing a clear security journey for clients, backed by factual evidence of the capability to orchestrate and oversee security measures. Sustaining innovations do not require technological complexity or sophistication to be applied, rather they generate improvements and refinements to existing cybersecurity technologies and practices. These innovations maintain and extend the life cycle of current cybersecurity approaches.

### b. Incremental Innovations

Incremental innovation refers to evolutionary changes made to meet new customer requirements or adapt to emerging technologies. This aspect introduces a reactive approach to changing needs, producing enhanced innovations to maintain competitiveness [21]. For example, responding to new security service requests from internal stakeholders or customers that may require improving processes to comply with specific regulations or adjusting practices to suit unique operational environments like factory settings with limited internet connectivity. The business value derived from incremental innovation comprises of a collective rise in capabilities maturity, ensuring that all elements of the organization advance in response to new challenges, and the establishment of criteria for assessing innovative technologies before adoption. Incremental innovations are small, evolutionary advancements that contribute to the overall robustness and effectiveness of cybersecurity measures. They can be minor tweaks or enhancements with some degree of technological complexity or sophistication, that cumulatively make significant differences.

*c. Disruptive Innovations*

Disruptive innovations introduce revolutionary changes, often appearing initially less adequate but eventually providing significant business value. It does not necessitate high costs or complex technology but focuses on unlocking new areas of customer engagement or technological application [20]. An example could be developing a new security control framework for ICS (Industrial Control Systems) or enabling service expansion to new business lines. Disruptive innovations also include adopting innovative technologies such as blockchain or artificial intelligence. The business value here lies in streamlining technical security with policymaking, thus providing clear understanding and empowerment in executing security roadmaps and preparing the organization for emerging cyber technologies and trends. Oftentimes breakthroughs happen in this aspect that fundamentally alter the landscape of digital security within an organization. These are often unexpected, coming from outside the traditional cybersecurity domain, and can completely change the rules of the game. For example, the adoption of blockchain technology to secure supply chain data, providing a tamper-proof and transparent ledger for tracking components and materials in the transportation sector. Moreover, the cross-collaborative way of workings in value chains can be seen as a disruptive innovation, initially. Mindset changes in the way of working for a team of cyber threat intelligence, maybe seen as a disruptive innovation. For instance, working every day holistically as a team split into different threat actor verticals, rather than having a designated person per day monitoring threat actors' activity horizontally. Such innovative ideas can be seen as disruptive in the beginning, nonetheless, over time the business value skyrockets, as the capabilities and teams more effectively and efficiently collaborate to increase cyber resilience while increasing their maturity at the same time.

*d. Transformative Innovations*

Transformative innovation represents a radical shift in how things are done, often leading to the substitution, or merging of capabilities or technologies. It requires a significant transformation, potentially necessitating a new skill base [23]. For instance, a transformative innovation could be exploring the use of AI in cybersecurity, or the convergence of AI with blockchain and cybersecurity. This type of innovation is about forward-thinking and thought leadership in adopting and utilizing emerging technologies. The business value enters by increased customer trust and brand reputation, as pioneering efforts in cybersecurity can provide a competitive edge and enhance overall cyber resilience. Proactively researching and implementing quantum-resistant cryptographic methods to prepare for the advent of quantum computing, radically altering the approach to data security, is another example of transformative innovation now. Furthermore, an AI-powered predictive maintenance mechanism is another example. Utilizing artificial intelligence to predictively analyse infrastructure health, identifying potential issues before they become critical, thus revolutionizing maintenance strategies. Or using an AI powered cyber threat intelligence pipeline that steers the cyber defences while providing automated cyber threat mitigations [20].

*2. Cybersecurity Innovation Value Proposition Score*

Business value in cybersecurity, particularly within the transportation and infrastructure industry, is multifaceted. Primarily revolves around the protection of assets and continuity of operations but it is also about the trust that users place in these critical systems. The value is derived from the effectiveness and efficiency of security measures, their alignment with the organization's strategy, and their contribution to the cyber resilience of the infrastructure. Thereby cybersecurity acts as a business enabler, rather than a showstopper to business objectives. To evaluate the business value of cybersecurity innovations, we propose two models that organisations may use subject to their maturity and expertise for innovation funnelling, namely, a semi-quantitative, and a fully quantitative. Both are multi-dimensional, yet basic models and not mutually exclusive. In fact, they could potentially work synergistically. The latter model considers the cost-benefit analysis, the impact on risk posture, and the enhancement of operational capabilities. As a result, we introduce three key formulas: Risk Reduction Value (RRV), Operational Efficiency Value (OEV), and Cost-Benefit Value (CBV). Each formula captures distinct dimensions of value, providing a quantitative basis for evaluating the efficacy of cybersecurity measures. The former model introduces two additional parameters that can be semi-quantitatively measured, namely, strategic alignment and trust.

*a. Semi-Quantitative model*

We begin by introducing the term 'Cybersecurity Innovation Value Proposition Score' (CIVPS) inspired by the work of Covin et al. [21], a compound index designed to evaluate ideas across six dimensions: revenue enhancement, cost efficiency, operational efficiency, risk mitigation, trust building potential, and strategic alignment. Each dimension is scored using a consistent scale ranging from 1 to 10, indicative of the estimated potential impact. This evaluation requires the consideration of multiple inputs by the cybersecurity innovation forum, involving a range of stakeholders. Consequently, the process of averaging the scores becomes imperative. The dimensions are depicted in figure 3.

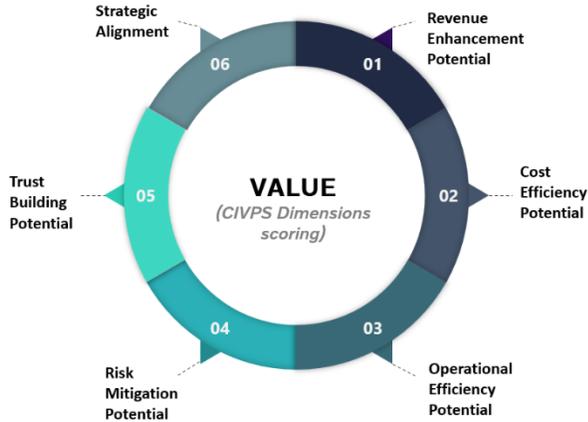

Figure 3 - Cybersecurity Innovation Value Proposition Score (CIVPS).

- **Revenue Enhancement Potential:** potentially innovative ideas are evaluated for their capacity to generate new financial inflows or augment existing revenue streams. This dimension can also be used to assess the potential for delivering value to stakeholders or fulfilling organizational missions or needs.
- **Cost Efficiency Potential:** measures an idea's ability to reduce current expenses, extend the life of existing assets, or pre-empt future expenditures, thus positively impacting the organization's cost structure.
- **Operational Efficiency Potential:** measures an idea's ability to streamline workflows, reduce/increase capability's quality deliverables, reduce "time-to-market", or eliminate non-value-adding activities within operational processes.
- **Risk Mitigation Potential:** ideas are scrutinized for their potential to address known vulnerabilities, enhance resilience, and reduce both the likelihood and impact of operational disruptions.
- **Trust Building Potential:** this dimension considers whether an idea can improve stakeholder perception, organizational perception, either externally or internally, fulfil or surpass customer expectations, and contribute to the organization's overall brand equity.
- **Strategic Alignment:** alignment with the organization's strategic direction provides a dual focus on both the intrinsic value and strategic fit.

Ultimately, scoring should not be the endpoint for all proposals. Often, ideas that do not pass the threshold of the dimensions in the early stages are not inherently deficient but may simply require more elaboration or maturation. These ideas, which may be premature due to the current state of technology or cost considerations, should be returned to their originators for further refinement. With adequate development and a more favourable technological context, these ideas could be reintroduced for consideration in future evaluation rounds.

*b. Quantitative model*

**Risk Reduction Value (RRV)** measures how much risk is mitigated by a cybersecurity innovation. This can be calculated by estimating the potential loss from cyber threats and the reduction in probability of these threats due to the innovation. $PL_{\text{before}}$ represents the potential financial losses due to cybersecurity threats prior to the implementation of a specific innovation. Similarly, let $PL_{\text{after}}$ denote the potential losses after the implementation, assuming a decrease due to the innovation's impact. We define $P_{\text{reduction}}$ as the probability reduction of a cybersecurity threat's occurrence as a result of the innovation. The RRV is then given by the equation:

$$RRV = (PL_{\text{before}} - PL_{\text{after}}) \times P_{\text{reduction}}$$

**Operational Efficiency Value (OEV)** quantifies the improvement in operational efficiency. It includes metrics such as reduced downtime or faster threat response times, subject to the context of capability measured. Let $G_{\text{operational}}$ denote the gains in operational efficiency that arise from the innovation, which include reductions in threat detection and response times or the increased automation of security processes, for instance. Let $C_{\text{implementation}}$ represent the total cost of implementing the innovation. Thus, the OEV is calculated as:

$$OEV = \frac{G_{\text{operational}} - C_{\text{implementation}}}{C_{\text{implementation}}}$$

This ratio defines the improvement in operational efficiency in relation to the implementation cost, offering an efficiency measure of the innovation's performance.

**Cost-Benefit Value (CBV)** assesses the cost savings against the investment in the cybersecurity measure. It considers both direct costs (e.g., implementation costs) and indirect costs (e.g., people upskilling / training costs or technology stack maintenance). $Total\_Savings$ aggregate all financial savings yielded by the innovation, including decreased losses from breaches, and improved operational efficiency. Conversely, $Total\_Costs$ aggregates all expenses associated with the innovation, incorporating initial outlay, maintenance, and any other related costs. The CBV is therefore calculated as:

$$CBV = \frac{Total\_Savings - Total\_Costs}{Total\_Costs}$$

This formula provides a holistic view of the financial benefits of the cybersecurity innovation against its total cost, summarizing the cost-effectiveness of the investment.

In many cases it is highly likely that innovations may introduce uncertainties. The use of Monte Carlo simulations in our framework provides for a thorough risk analysis and a probabilistic understanding of business value, which is critical for making strategic decisions under uncertainty [22]. For instance, assuming we are evaluating the potential cost savings from preventing cyber incidents over a given period through an innovative blockchain based intrusion prevention system. Let $C_{incident}$ represent the potential costs of cybersecurity incident without the investment on the innovative system. Let $P_{incident}$ be the probability of such an incident occurring within a specific period. Let $C_{investment}$ be the cost of the investment. Let $R_{investment}$ be the reduction in the probability of the incident due to the investment. We run *N* iterations, where in in each iteration *i*, we simulate whether an incident occurs based on $P_{incident}$ and $R_{investment}$ and calculcate the cost savings if the incident is prevented. Next, we calculate the average expected savings across all iterations to estimate the business value of the investment.

$$BV_{cyber} = \frac{1}{N}\sum_{i=1}^{N}\bigl(C_{incident} \times I_{prevented,i} - C_{investment}\bigr)$$

Let $BV_{cyber}$ represent the estimated business value of the investment. *N* is the number of iterations in the simulation. Let $C_{incident}$ be the cost of cybersecurity incident. Let $I_{prevented,i}$ represent a binary indicator (0 or 1) for whether an incident is prevented in iteration, and *I* determined stochastically based on $P_{incident}$ and $R_{investment}$. Let $C_{investment}$ be the cost of the investment.

In each iteration, $I_{prevented,i}$ s determined by generating a random number and comparing it to the adjusted probability of an incident due to the investment. If the random number is lower than $P_{incident} \times (1-R_{investment})$ then $I_{prevented,i}$ is set to 1, indicating that an incident would have occurred but was prevented due to the investment. The cost savings for that iteration are then $C_{incident}$ minus $C_{investment}$ By averaging these savings over *N* iterations, we obtain an estimate of the investment's business value.

This Monte Carlo simulation approach provides for a detailed analysis of the uncertainty and variability associated with cybersecurity risks and the potential savings from investments on innovations, thus, ultimately guide strategic decisions through a quantifiable manner. Together, these formulas form the analytical backbone of the quantitative model, presenting a method for the quantitative evaluation of cybersecurity innovations within this framework.

### *3. Balance Resource and Risks*

In this phase we estimate the necessary effort for implementation in relation to the expected impact. We also estimate the total investment and effort required for an idea to be scaled and adopted within the organization. This assessment is crucial as certain ideas, while potentially straightforward during conceptualization and validation, may need considerable time, resources, or organizational disruption upon deployment. Early recognition of these factors is imperative to establish that the full scope of the resource commitment is clear to all stakeholders. Typically, in this phase, and oftentimes in the earlier phase (2), proposals undergo review for approval and seek endorsement from senior management.

#### *a. Required Effort Estimation*

To estimate the required effort, it is recommended to formulate estimations in response to queries such as: what quantity of expertise, time, and financial resources are needed to diminish the uncertainties surrounding the concept and to finalize a proof of value (PoV)? What challenges are anticipated in integrating the new concept with existing systems or processes? Are there regulatory approvals or compliance standards that the concept must meet? How extensive is the stakeholder engagement process expected to be? Additionally, what are the projected financial prerequisites for the concept to be embraced organization-wide?

Focusing solely on the initial financial costs needed to investigate an innovative prospect may yield a biased anticipation of the subsequent expansion and integration process. This has the potential to dismiss scenarios where a concept is rapidly validated within weeks, yet the scaling and organizational adoption may span years, demanding substantial financial and manpower investments to fully realize the idea on a larger scale.

#### *b. Implementation Impact Estimation*

To estimate the implementation impact, it is critical to evaluate whether the idea requires a comprehensive shift in the organizational structure or if its implementation is more localized, thereby response is required to queries such as: does the implementation of the idea need a comprehensive organizational transformation, or is the scope of adoption more limited? Are there requirements for the establishment of new operational processes, governance frameworks, or reporting mechanisms? To what extent will the organization need to modify or upgrade its existing technological infrastructure to accommodate the innovation?

Eventually this is a high-level subjective assessment. Nonetheless, it communicates to the larger organization the attention given to future implications in the innovation process. It also underscores the idea that the quick and iterative pace of innovation does not neglect the assessment of its long-term strategic effects. The results of the assessment may be effectively depicted using a straightforward XY axis graph, as illustrated in figure 4.

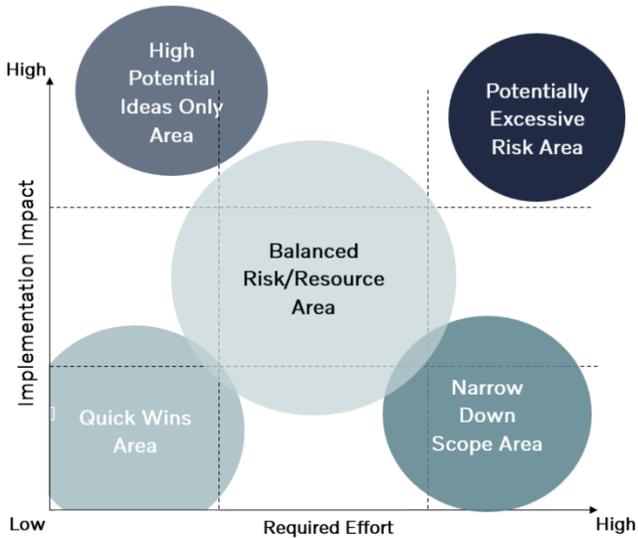

*Figure 4 – Cybersecurity innovation ideas road mapping.*

Innovative ideas positioned in the quadrant of low required effort and low implementation impact typically represent straightforward, achievable targets, known as quick wins. These can be progressed to execution with even moderate initial CIVPS. Conversely, ideas situated within the quadrant requiring high effort and high implementation impact are considered ventures with substantial risk. Such initiatives are advisable only if they show exceptional potential and are among the top-tier disruptive or transformative ideas. Ideas that demand considerable effort but are expected to have a minimal implementation impact call for a scope reassessment prior execution. The strategy should be to find a more contained scope that minimizes the associated risks and supports quicker value realization iteratively and agile. Lastly, ideas requiring minimal effort but offer a significant implementation impact, a "go" signal for execution should be given under conditions upon their extraordinary potential.

*4. Execution and Value Realization*

In this last stage, the focus shifts to translating vetted cybersecurity innovations into tangible outcomes for the organization. The stage begins with establishing specific timelines, milestones, budget, and risk management strategies following up from the previous stage's outcomes. Next, a project team is formed with clearly defined roles and responsibilities. Typically, a prototype development is crucial, or a minimum viable product (MVP) is essential to test the ideas in real-world scenarios [28]. Several other terminologies and concepts apply at this stage, such as proof of concept (PoC) or proof of value (PoV), subject to organizational needs and dynamics. Moreover, testing and validation alongside stakeholder management naturally should happen at this stage. The team is required to have regular communication with all stakeholders for feedback and alignment while testing and validating the prototype or MVP. The Cybersecurity innovation forum holds a crucial role throughout all stages including execution, providing ongoing support to ensure the innovative solutions are effectively integrated, stakeholders are fully engaged, and the value is ultimately realized. This stage is intentionally designed to be modular, thereby allowing for a high degree of flexibility and customization according to organizational needs and project management methodologies. This design choice enables the framework to accommodate a wide range of innovation execution scenarios, ultimately allowing the outcomes to be both effective and closely aligned with organizational objectives.

IV. DISCUSSION

The adoption of a business value-driven framework for cybersecurity innovation represents a paradigm shift for organizations in the transportation and infrastructure sectors. Cyber-foresight enables the strategic alignment of cybersecurity initiatives with business objectives, thereby organizations can better justify investments in cybersecurity, align initiatives with broader strategic goals, ultimately increase stakeholder confidence.

This approach diverges from conventional market-driven strategies, where oftentimes prioritize broad applicability and the potential for commercialization. Such models have driven substantial technological advancements, nonetheless, they may not always address the specific needs of critical infrastructure sectors or dedicated cybersecurity innovation capabilities. Our framework, by contrast, provides a more nuanced approach, where the value of innovation is measured by its direct impact on operational and cost efficiency, risk mitigation, strategic alignment, compliance with sector-specific regulations and brand building equity. This targeted approach is particularly beneficial in sectors where cybersecurity is integral to operational continuity and public safety.

However, certain limitations must be acknowledged. Smaller organizations may face challenges in allocating sufficient resources for disruptive and transformative innovations. Moreover, while we empirically suggested an optimal sector-specific balance through innovation categorization, striking the right balance can be complex and requires ongoing adjustment. Finally, the rapid pace of technological advancements in cybersecurity may require regular reassessment of the framework's relevance.

V. CONCLUSIONS & FUTURE RESEARCH

In this paper, we presented a cybersecurity innovation framework for the transportation and infrastructure sectors based on business value generation rather than one-size-fits-all market driven approach. We provided a structured method for organizations to critically assess and prioritize their cybersecurity initiatives, enable cyber-foresight, help innovative ideas to contribute to the overall strategic goals, eventually enhancing the cyber resilience. The importance on sustaining innovation promotes systematic innovation within teams, encouraging a culture of continuous improvement. The quantitative and semi-quantitative options to measure value provide a data-driven evaluation of cybersecurity initiatives, which aligns them with strategic business objectives and ultimately enhances decision making. Lastly, the strategic focus on transformative and disruptive innovations assists organizations to proactively address emerging cybersecurity threats and adapt to the evolving cyber threat landscape. Future work is needed to refine these models while exploring the application of this framework in different contexts and scales, particularly in smaller organizations where the resource capacity is an inherent challenge.

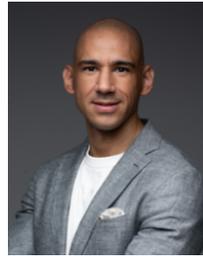
Lampis Alevizos received his M.Sc. degree in cybersecurity and his Ph.D. degree in Computer Science from the University of Central Lancashire (UCLan), with research focused on the convergence of ZTA, blockchain, and DLT with cybersecurity. Lampis holds and actively maintains several industry certifications, including CISSP, CCSP, CCSK, CISA, CISM, among others. He is currently the Head of Cyber Defence Innovation at Volvo Group.

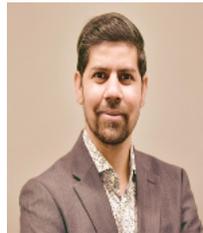
Lalit Bhakuni earned his M.Sc. degree in Computer Science from SMU India and an MBA from the National Institute of Management India. He has earned and continues to uphold numerous esteemed industry certifications such as CISSP, OSCP, CISM and GCFA, among others. Currently, Lalit serves as a Vice President at Volvo Group where he leads the Active Cyber Defence Organization.

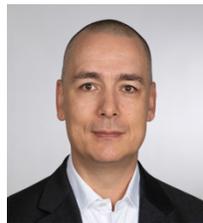
Stefan Jäschke is a Senior Vice President at Volvo Group, heading the Enterprise IT Security function. He and his team build, operate and manage group-wide cybersecurity capabilities. He is an advocate of threat intelligence-driven cyber defence and focuses on aligning business strategy with cybersecurity innovations to enable Volvo Group's digital transformation.